\documentclass[apjl]{emulateapj}

\shorttitle{The Lithium-Rich Red Giant Rapid Rotator G0928+73.2600} %Adding G0928z+73.2600 makes it 45.. max is 44 :-P
\shortauthors{Carlberg et al.}

\begin{document}

\title{The Super Lithium-Rich Red Giant Rapid Rotator G0928+73.2600: 
A Case \\ for  Planet Accretion?}

\author{
Joleen K. Carlberg\altaffilmark{1},
Verne V. Smith\altaffilmark{2}, 
Katia Cunha\altaffilmark{2,3,4},  
Steven R. Majewski\altaffilmark{1}, and
Robert T. Rood\altaffilmark{1}}

\altaffiltext{1}{Department of Astronomy, University of Virginia, P. O. Box
400325, Charlottesville, VA 22904-4325, USA ; jkm9n@virginia.edu}%(jkm9n, srm4n,rtr@virginia.edu)}
\altaffiltext{2}{National Optical Astronomy Observatory, 950 North Cherry Avenue,  Tucson, AZ 85719, USA}
\altaffiltext{3}{Steward Observatory, 933 North Cherry Avenue,  Tucson, AZ 85121, USA}
\altaffiltext{4}{on leave from Observat\a'{o}rio Nacional, R. Gal. Jos\a'{e} Cristino, 77, 20921-400
S\~{a}o Crist\a'{o}v\~{a}o, Rio de Janeiro, RJ, Brazil
}

\begin{abstract}
We present the discovery of a super lithium-rich K giant star, G0928+73.2600.
This  red giant ($T_{\rm eff} =4885$~K  and $\log g=2.65$) is a fast rotator with
a projected rotational velocity of 8.4 km~s$^{-1}$ and an unusually high lithium abundance of {\it A}(Li)~$=3.30$ dex. 
Although the lack of a measured parallax precludes knowing the exact evolutionary phase, an isochrone-derived estimate of its luminosity places the star on the
 Hertzsprung--Russell diagram in a location that is {\it not} consistent with either the red bump on the first ascent of the red giant branch or with the second ascent on the asymptotic giant branch,  the two evolutionary stages where lithium-rich giant stars tend  to cluster.  Thus, even among the already unusual group of lithium-rich giant stars, G0928+73.2600 \ is peculiar.
 Using $^{12}$C/$^{13}$C \ as a tracer for mixing---more mixing leads to lower $^{12}$C/$^{13}$C---we find $^{12}$C/$^{13}$C~$= 28$, which  is near the expected value for standard first dredge-up mixing.  We can therefore
 conclude that  ``extra''  deep mixing has not occurred.   Regardless of  the ambiguity of the evolutionary stage,  the extremely large lithium abundance and the rotational velocity of this star are unusual, and we speculate that G0928+73.2600    has been enriched in both lithium and angular momentum from a sub-stellar companion.

\end{abstract}


\section{Introduction}
\begin{deluxetable*}{cccccccc}[t]
\tablewidth{0pc}
\tablecaption{Derived Stellar Parameters and Abundances\label{stellparam}}
\tablehead{
  \colhead{Obs. Date} &
  \colhead{$T_{\rm eff}$} &
  \colhead{log $g$} &
  \colhead {[Fe/H]} &
  \colhead{$\xi$} &
  \colhead{$v \sin i$} &
  \colhead{{\it A}(Li)$_{\rm LTE}$}  &
    \colhead{{\it A}(Li)$_{\rm NLTE}$}  \\
(yyyy-mm-dd)	    & (K) & (dex) & (dex) & (km~s$^{-1}$)& (km~s$^{-1}$) & (dex) & (dex)}
\startdata
2007-03-06 & 4900 & 2.7 &  -0.26 &1.51  &8.4& 3.62 & 3.296\\  
%NLTE COR = -0.324 = 3.296
2008-01-11 & 4870 & 2.6 & -0.23 & 1.41 & ... & 3.61 & 3.308 \\ 
%NLTE COR =-0.302  = 3.308
\hline
Mean & 4885 & 2.65 & -0.245 & 1.46 & 8.4 & 3.62 & 3.30
\enddata
\end{deluxetable*}

The expected abundance of lithium, {\it A}(Li)\footnote{{\it A}(Li)=$\log(N_{\rm Li}/N_{\rm H})+12$.}, 
in the stellar atmosphere of a red giant star depends on many factors, including the stellar mass and the current evolutionary stage. % evolutionary stage of the star.
It is well understood that lithium should be destroyed as the deepening convective layers of evolving red giants mix the lithium into the hot interior of the star, diluting the surface abundance \citep{iben67}. 
 Depleted  lithium is then expected throughout most of the red giant phase.   In stars more massive than 1.5~$M_{\sun}$ and more
 luminous than $\sim10^4L$~$_{\sun}$, 
 temperatures at the base of the convection envelopes are hot enough for the nucleosynthesis of  $^7$Li  through the Cameron--Fowler chain,  and if the convective mixing time is faster than the lithium destruction time, an abundance of lithium can be built up in the envelope \citep{scalo75}. Stars meeting these requirements are generally late in the second ascent, i.e.,  the asymptotic giant branch (AGB) stage.

 Contradicting these theoretical expectations  are the many red giant stars that  show {\it A}(Li) values that either exceed or fall short of the anticipated value. First-ascent, low-mass stars, for example, generally show far more lithium depletion than the standard dilution models \citep{brown89}; the {\it A}(Li) in these stars point towards additional mixing processes beyond the standard model predictions.
Even more surprising are the stars for which {\it A}(Li) exceeds not only the standard dilution model predictions, but also the upper limit of primordial abundance measured from the ISM/meteorites in our solar 
system---{\it A}(Li)~$= 3.28$ \citep{lodders09}.  These stars are difficult to account for, especially stars on the red giant branch (RGB), which have  convection
envelopes that are too cool to regenerate lithium.  A special mixing mechanism, termed ``cool-bottom processing'' \citep[CBP,][]{sack99} %,nollett03}
 is needed both to get material below the convection zone to layers hot enough for the Cameron--Fowler process to work and then to transport lithium back into the convection zone where regular convective mixing can rapidly distribute fresh lithium throughout the stellar envelope. 
  
  Clues to identifying the physical mechanism responsible for CBP can be found by looking at the properties of stars that show unusual lithium enhancements.
 \cite{charbonnel00}, hereafter C00,  and \cite{reddy05} noted that super lithium-rich giants tend to cluster in the Hertzsprung--Russell (H-R) diagram in two groups.
   The first group is near the red bump (or luminosity bump), which is an evolutionary stage  of low-mass stars on the RGB when the outward hydrogen-burning shell reaches the chemical discontinuity left behind by the convection zone at the peak depth of first dredge-up.  %\cite{charbonnel00}
C00 hypothesized that in this phase, low-mass stars may go through a short-lived burst of lithium production that is quickly diluted. 
 The other lithium-rich group seen in the C00 study is on the early  AGB, at $\log L/L_{\sun}\sim 2.8$.  These AGB stars are more massive and did not go through a red bump
 stage while on the RGB.

\cite{sack99} studied the effects of CBP on the surface abundances of light elements by including parameterized models of CBP in red giant stars at the red bump. They found that, depending on 
the mixing geometries, CBP can explain both the general destruction of light elements and the occasional creation of $^7{\rm Li}$.  However, they did not provide a physical mechanism behind their
parameterized model.   Rotation was thought to be a likely candidate mechanism until \cite{palacios06} found that a self-consistent model of rotational mixing could not generate enough circulation 
to be responsible for CBP.
A recent paper by \cite{palmerini10} summarized various mechanisms that have been explored as the physics of CBP and noted that the two current contenders  are
thermohaline mixing \citep{eggleton06,charbonnel07} and magnetic buoyancy  \citep{guandalini09}.
Because thermohaline mixing is a relatively slow process, it is generally invoked only to explain additional lithium {\it depletion}; the mixing may be too slow to bring fresh lithium to the stellar 
envelope. The magnetic buoyancy models are fast enough and can work both at the red bump and on the AGB; however,   both the \cite{guandalini09} and \cite{palmerini10} models 
have maximum lithium enrichments of {\it A}(Li)~$\sim$~ 2.5 dex---well short of the lithium abundances observed in the most lithium-rich red giants. 

A useful tool for tracking the amount of mixing in  a star that may have regenerated lithium is $^{12}$C/$^{13}$C because both lithium and $^{12}$C/$^{13}$C are reduced during mixing episodes  in  the absence of the Cameron--Fowler mechanism.   Consequently, small values of $^{12}$C/$^{13}$C are expected when extra mixing processes  succeed in replenishing lithium in red giant atmospheres.  In light of this expectation, the most
unusual stars are those with relatively large values of $^{12}$C/$^{13}$C, which suggests {\it  standard mixing}, and super Li-rich abundances (near or above the meteoritic value) suggestive of lithium regeneration, which  requires {\it extra mixing}.  

While much theoretical work on mixing processes  focuses on trying to understand the two groups of lithium-rich giants identified in C00,  it is worth noting that dividing the lithium-rich giants into two categories may still be too simple of a picture.   A recent review of lithium in red giants by \cite{smith10}  highlights examples  of lithium-rich giants at many different phases along the RGB;  many physical processes may be contributing to the population of lithium-rich giants.

In this Letter, we announce a discovery of just such an unusual lithium-rich star, G0928+73.2600, which was originally selected from the  Grid Giant Star Survey \citep{ricky01} for a spectroscopic survey of slow and rapid rotator RGB stars collected for chemical abundance studies.  This star  has {\it A}(Li)~$=3.30$~dex and $^{12}$C/$^{13}$C~$\sim 28$, and it does not fall into either of the lithium-rich groups identified in C00.
The star  also  has enhanced rotation, with $v \sin i$~$=8.4$~km~s$^{-1}$.
We place this star in context of the other known lithium-rich giants and lithium-regeneration mechanisms.  In Section 2, we describe the observations and stellar parameter/abundance analysis of G0928+73.2600.  Section~3 provides a comparison of  the evolutionary phase of G0928+73.2600 to other known Li-rich stars and discusses implications for the mechanism responsible for the excess lithium.
Our conclusions are presented in Section 4.

\section{Observations and Abundance Measurements}
 
 Two high-resolution spectra of G0928+73.2600 were obtained with  the echelle spectrograph on the Kitt Peak Mayall telescope, at a signal-to-noise ratio (S/N)~$> 100$ per pixel and $R \sim43,000$.
The spectra  were reduced using standard IRAF procedures and the echelle orders combined and continuum corrected to 
create two normalized one-dimensional spectra. 

Stellar abundances were derived using the MOOG stellar line analysis program 
\citep{sneden73} and MARCS spherical stellar atmosphere models \citep{marcs08}.  A set of 73 \ion{Fe}{1} lines  constrained
the effective temperature and microturbulence at the point where there is no trend of iron abundance with either excitation potential or reduced equivalent width.  Thirteen \ion{Fe}{2} lines
constrained surface gravity.    The line list was compiled from a variety of sources including the Vienna Atomic Line Database 
\citep{vald} and \cite{mandell04}.
Table \ref{stellparam} shows the stellar parameters derived from each spectrum.
The average solution for G0928+73.2600 is $T_{\rm eff}=4885\pm30$~K, $\log g=2.65\pm0.1$~dex, [Fe/H]~$= -0.25\pm0.03$~dex, 
and $\xi=1.46\pm0.10$~km~s$^{-1}$.  

The rotational broadening was derived from the \ion{Fe}{1} line at 6750.15~\AA, which was chosen because it is in the same spectral order 
as the \ion{Li}{1} line and is free from blending.  Instrumental broadening was measured from the ThAr spectrum. 
The macroturbulent broadening, $\zeta=5.62$~km~s$^{-1}$, comes from the temperature relation of \cite{Hekker07} for class III giants.  With $\zeta $ and instrumental broadening fixed, 
we used a $\chi^2$-minimization routine   to find the rotationally broadened  synthetic spectrum that fit best, yielding $v \sin i$~=~8.4~km~s$^{-1}$ for G0928+73.2600.

\begin{figure}[t]
\begin{center}
\includegraphics[scale=0.35]{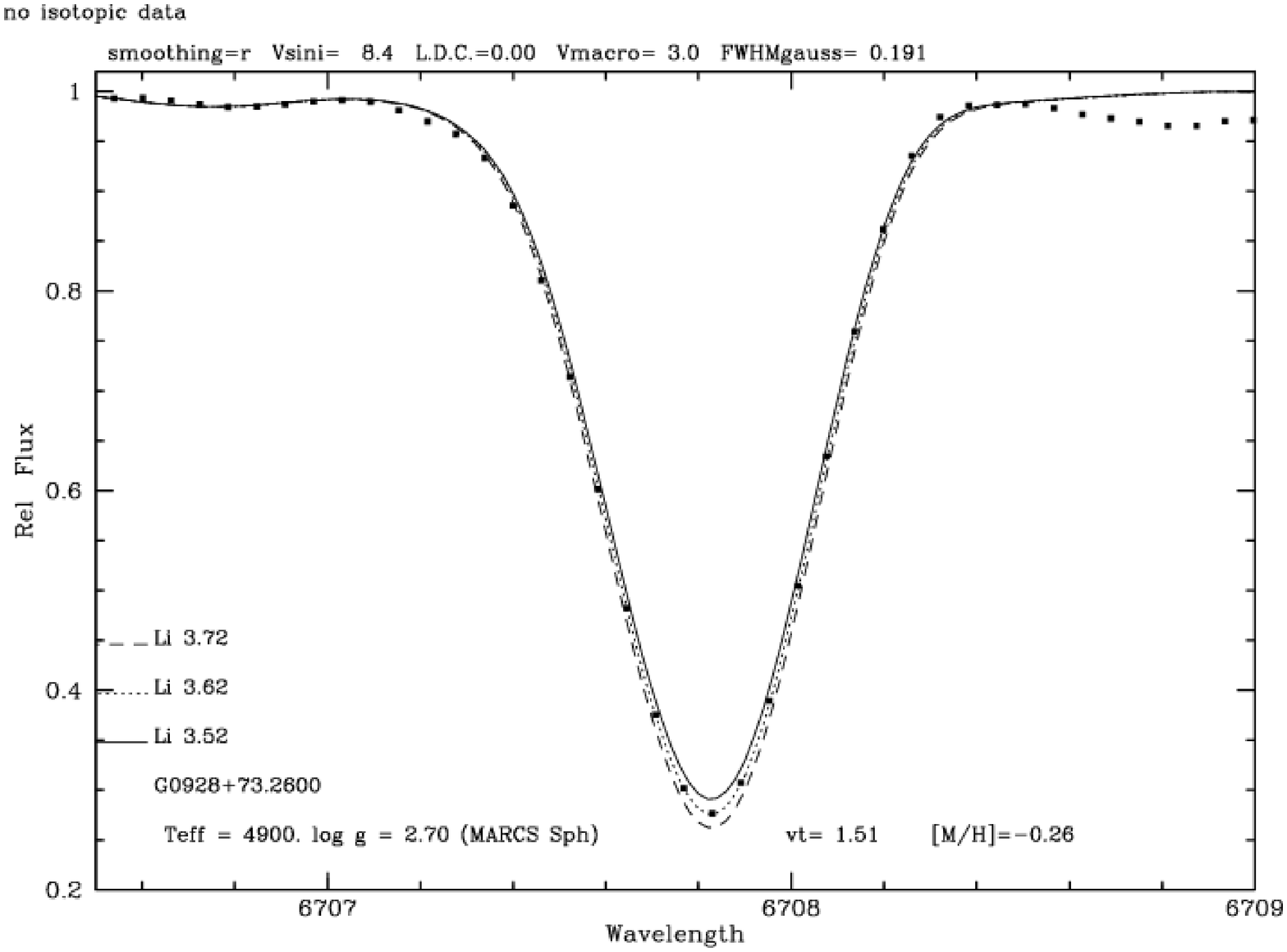}
\caption{ Fit from MOOG  to the lithium resonance lines at 6707.8--6707.9~\AA\ in the 2007 spectrum.
 The data are the small circles, and the lines show the best-fit {\it A}(Li) (dotted), and 0.1 dex above (dashed) and below (solid)
the best fit. \label{fig:moog}}
\end{center}
\end{figure}

Finally, to measure the lithium abundance, we used spectral synthesis to fit the spectral region around the Li line region at 6707~\AA, as illustrated in Figure \ref{fig:moog} using the line list published in \cite{ghezzi09}.  Free parameters  in the fit include {\it A}(Li) and small adjustments in the overall continuum level, velocity solution, and broadening to get the best fit.
From this analysis, we find an LTE solution of {\it A}(Li)~$=3.62\pm 0.07$~dex for G0928+73.2600, which required reducing $\zeta$ to 3.0~km~s$^{-1}$.  
The quoted error in {\it A}(Li) includes contributions from both the fitting procedure and variations within the errors of the stellar parameters.
The latter were computed by holding the equivalent width associated with the lithium abundance constant and adjusting the stellar parameters within the error bars to see how {\it A}(Li) varied. We found that
temperature introduced the largest error at 0.04~dex.   Finally, we computed non-LTE corrections to the lithium abundance by interpolating the \cite{lind09} grid of corrections to our stellar parameters. These corrections yield a non-LTE Li abundance of 3.30~dex.

\begin{figure}[h]
\begin{center}
\includegraphics[scale=0.38]{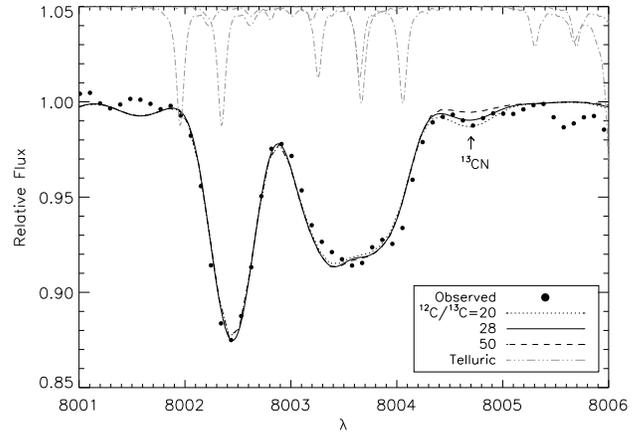}
\caption{ Fit to CN lines in the G0928+73.2600 spectrum. The observed data, a sum of the 2007 and 2008 spectra, are plotted as circles.  The lines show synthetic spectra for carbon ratios of 28 (solid, the best fit), 20 (dotted) and 50 (dashed).  The gray dash-dot lines show an atlas telluric spectrum shifted in velocity to the stellar rest frame.  Note that the $^{13}$CN lines are free of telluric contamination. \label{fig:cratio}}
\end{center}
\end{figure}

We measured $^{12}$C/$^{13}$C  using spectral synthesis of the CN features between 8001 and 8006~\AA.   Because of the lower S/N in this part of the spectrum, we added
the  two observations together to increase the S/N. Our fitting routine allows for   
variations in  the carbon ratio, C and N abundances (keeping C/N fixed at 1.5), velocity, and  overall scaling.  Figure \ref{fig:cratio} shows the summed spectrum of G0928+73.2600. 
 The lines between 8003 and 8004.3~\AA\ are $^{12}$CN features, whereas  $^{13}$CN   forms the lines near  8004.7 \AA.  
We measure a best-fit $^{12}$C/$^{13}$C \  of $28\pm 8$ from this spectrum.    The  weak $^{13}$CN lines are responsible for the large error bars. 

\section{Discussion}
As mentioned earlier, $^{12}$C/$^{13}$C can be used as a tracer for mixing in giant stars.
Early stellar models predict post first dredge-up carbon ratio values near 23 \citep{sweigart89}. More recent  models of mixing in red giant stars indicate that post first dredge-up values of  $^{12}$C/$^{13}$C for 1 and 2 $M_{\sun}$ stars range from 29.5 to 22.3, respectively \citep{eggleton08}.  Empirical measurements verify these models;   \cite{gilroy91} found $^{12}$C/$^{13}$C~$\sim 22$ for giants at the end of first dredge-up, and even lower values near the RGB tip.
 Therefore, G0928+73.2600 has a measured $^{12}$C/$^{13}$C that is comparable, though slightly higher than  model predictions and empirical values for giants, suggesting that this particular star may not have completed the first dredge-up.  However, we note that the error bars make it difficult to be certain of these conclusions. The carbon ratio is certainly not very low.

\begin{figure}
\begin{centering}
\includegraphics[scale=0.38]{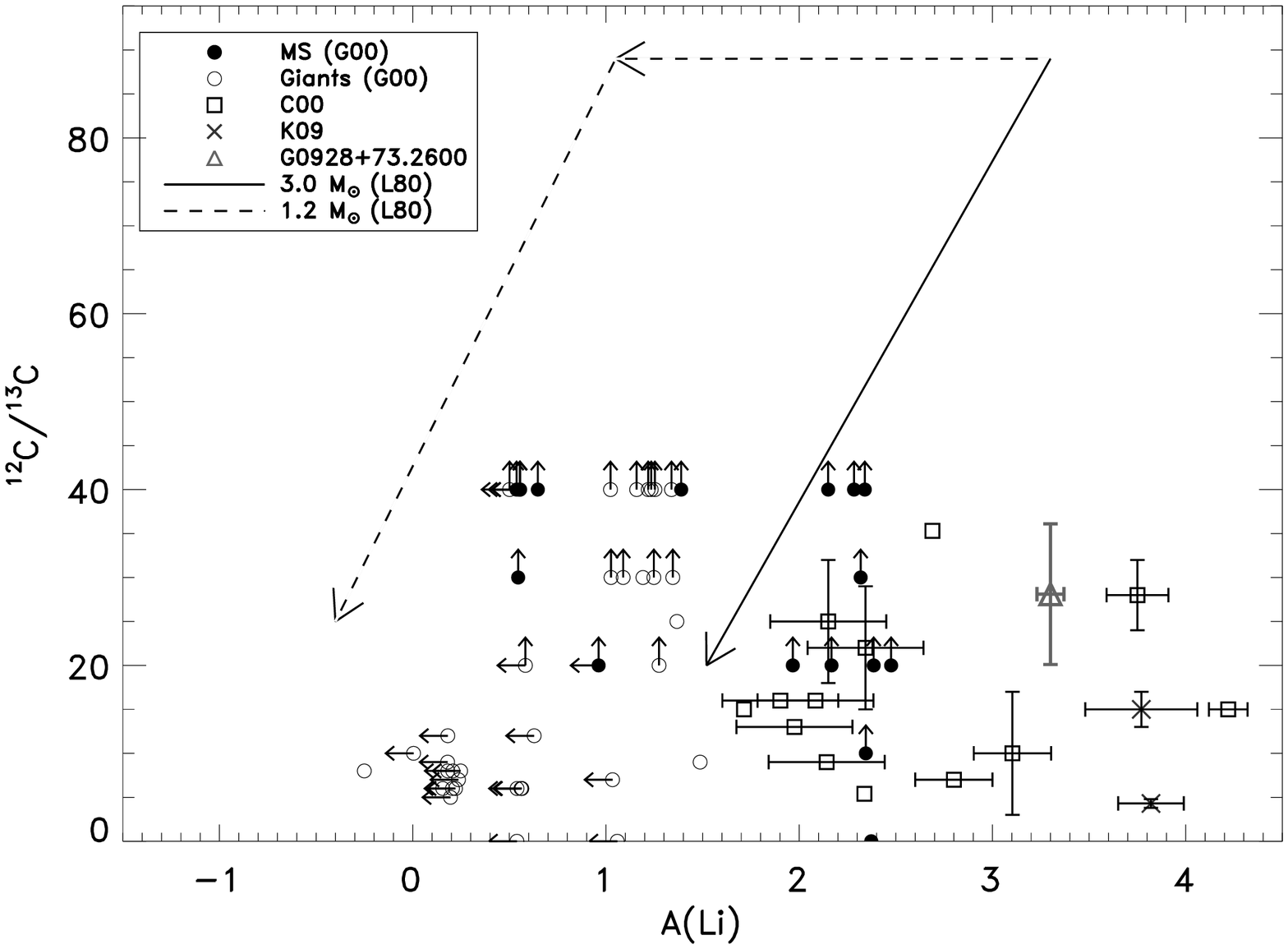}
\caption{Lithium abundances  and $^{12}$C/$^{13}$C ratios for stars with data in the literature. All open symbols denote giant stars, while filled symbols are main sequence stars.  The two models of evolving surface abundances are adapted from \cite{lambert80}.  The literature references in the legend are ``G00'' \citep{gratton00}, ``C00'' \citep{charbonnel00}, ``K09'' \citep{kumar09}, and ``L80'' \citep{lambert80}.  Error bars and limits are plotted when known.  The LTE results from the literature were corrected for departures from LTE using the non-LTE corrections in
 \cite{lind09}. \label{fig:ali_cratio}} 
\end{centering}
\end{figure}

How unusual is the combination of both high lithium and a $^{12}$C/$^{13}$C of 28?  To answer this question, we plot literature values of red giant stars (and some main sequence stars)  in Figure \ref{fig:ali_cratio}.  To properly compare these data to our own, we remove any previous NTLE corrections made to the lithium measurements  and apply the \cite{lind09} corrections ourselves.  We also plot two standard models of lithium dilution  and decreasing $^{12}$C/$^{13}$C in RGB evolution; these models are adapted from Figure 9 of  \cite{lambert80}. Most of the giant stars fall in the lithium-carbon ratio  space delimited by the two models.  For lithium levels exceeding the primordial abundance of 3.28 dex, all but two stars have $^{12}$C/$^{13}$C lower than 20: G0928+73.2600 and HD 9746.

\begin{figure}[t]
\begin{centering}
\includegraphics[scale=0.38]{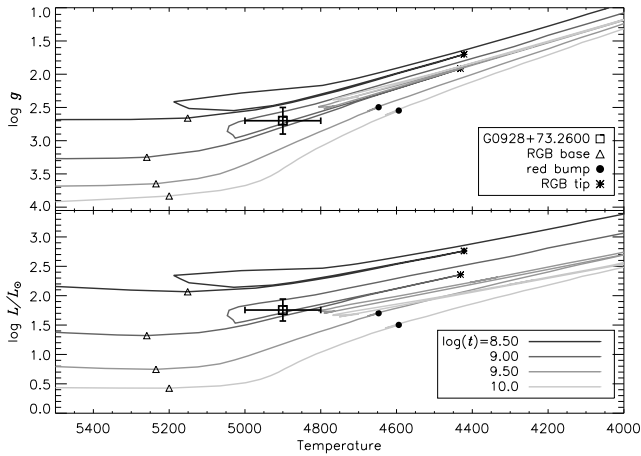}
\caption{Isochrones for $Z=0.012$ ([Fe/H]~$= -0.20$) for four different ages.  Major evolutionary phases are marked including the RGB base (triangle), RGB tip (*), and the red bump (circle). 
 G0928+73.2600 is plotted as a square.  The top panel shows temperature and surface gravity, while the bottom panel shows temperature and luminosity. 
The adopted luminosity range for G0928+73.2600 is the maximum and minimum luminosities in the isochrones covering the same temperature and gravity space enclosed by the error bars in the top panel.  The stellar mass of the isochrones within the error bars of G0928+73.2600 ranges from 2.0 to 2.2$M_{\sun}$.\label{fig:iso}} 
\end{centering}
\end{figure}

To gain a deeper understanding of the evolutionary stage of G0928+73.2600, we plot its stellar parameters together with \cite{giard00} isochrones in Figure \ref{fig:iso}.
Of the isochrones plotted, the one most consistent with the location of G0928+73.2600 is $\log t=9$; note that this
isochrone has no red bump (marked by the filled circles) because the stars are of relatively higher mass than the older isochrones.  Because of the error bars on G0928+73.2600's stellar parameters,  it is uncertain whether  the star is on the RGB, horizontal branch, or AGB.  Nevertheless,  we can use the isochrones to estimate a reasonable range of luminosity despite not knowing the distance.
The range of stellar masses and luminosities overlapping the temperature and gravity of G0928+73.2600 in the upper panel of Figure \ref{fig:iso} are 2.0-2.2~$M_{\sun}$ and $\log L/L_{\sun}\sim$~1.6-1.9.  Using these estimates, we plot 
G0928+73.2600 on the HR diagram in the bottom panel of Figure \ref{fig:iso} together with the same isochrones.  For stars in the estimated mass range of G0928+73.2600, lithium dilution is expected to begin at $T_{\rm eff}\gtrsim 5000$~K, first dredge-up ends at $\sim4600$~K, and the red bump for stars just below our estimated mass range occurs at $\sim4400$~K  (see C00, particularly Figure 1).
Consequently, G0928+73.2600 is an oddity; it is cool enough that it should be well into its lithium dilution stage, yet it is either too massive to even
evolve through the red bump or too hot to have reached the red bump if it is less massive than our estimate.
 The star's $^{12}$C/$^{13}$C combined with the mass estimate suggests that G0928+73.2600 has not yet completed first dredge-up---\cite{eggleton08} predict $^{12}$C/$^{13}$C~$=22.3$ for a 2$M_{\sun}$ star.  This fact places the likely evolutionary stage on the first ascent RGB star before the red bump. 

Recalling HD 9746, the other star sharing a  similar position in Figure \ref{fig:ali_cratio} as G0928+73.2600, we find that G0928+73.2600 is again the odd-star out.  HD 9746 is listed in C00 as one of the red bump lithium-rich stars, whereas G0928+73.2600 cannot be part of this group.   We must therefore consider anew the possible explanations for enhanced lithium in giant stars.
We already know that G0928+73.2600 should be well into the phase of  lithium dilution, yet the lithium abundance is near its main sequence value  \citep[see, for example, Figure 6 of][]{luck06}.   

Generally, large lithium abundances in giant stars can be explained by suppressed lithium dilution, lithium regeneration, and lithium replenishment.
The first scenario is ruled out by the $^{12}$C/$^{13}$C; mixing on the level of standard first dredge-up {\it has} occurred, i.e., $^{12}$C/$^{13}$C$\ll89$, the main sequence value.  Lithium regeneration has been explored by \cite{sack99}; their
model of parameterized CBP can regenerate lithium at this level.  
Figure 9  of their paper shows predictions for {\it A}(Li) as a function of RGB luminosity for a number of their models; however, lithium enhancements of the level seen in G0928+73.2600 occur in their models for only the most
luminous RGB stars---higher than that of G0928+73.2600.  Admittedly, their predicted luminosities for the lithium-rich stars also exceed that of  observed stars in C00.
The magnetic buoyancy  models of both \cite{guandalini09} and \cite{palmerini10}  can also regenerate lithium, but their models predict a maximum {\it A}(Li) of about 2.5 dex ---much lower than what is found in G0928+73.2600.

We also note that G0928+73.2600 is rotating  rapidly for a giant star,  with $v \sin i$ = 8.4~km~s$^{-1}$  when most stars of this type generally rotate with $v \sin i$~$<2$~km~s$^{-1}$\  \citep{deMed96b}.  
This enhanced rotation may be relevant to the lithium-regeneration models if it can create the favorable circumstances needed in the \cite{sack99} model by, for example, increasing the mixing speeds. 
 On the other hand, recall that \cite{palacios06} found that rotational mixing is not efficient enough to create lithium-rich stars.
 In either case, one must consider the origin of the excess angular momentum implied by the moderately fast rotation.  
 
 We speculate that the increased  angular momentum and the trigger for lithium regeneration could be related to the accretion of a planet.  \cite{siess99} computed detailed models of the effects of planets falling into their host stars and found that, if anything, lithium is depleted even more rapidly.  However, Siess \& Livio admit the possibility that planet accretion could trigger the \cite{delareza96} model, where lithium regeneration is associated with a period of mass loss and ejection  of a circumstellar shell.
 
 The planet accretion line of reasoning naturally leads to the third scenario of creating lithium-rich giant stars: lithium replenishment. 
Planet accretion was first put forward by \cite{alexander67} to explain lithium enrichment in giant stars and has been invoked by many authors since then to explain both lithium enhancements and rapid rotation in giant stars \citep[see, e.g, ][]
{wallerstein82,reddy02,drake02,carney03,denissenkov04}.   On the main sequence, \cite{israelian01} found evidence of accreted planetary material in the detection of $^6$Li in HD82943 \citep[although other authors, e.g.,][cannot reproduce the detection]{reddy02b,ghezzi09}.

G0928+73.2600's evolutionary status is consistent with the models of \cite{carlberg09} that predicted that enhanced rotation from planet accretion is most likely found on the lower RGB.
The rotation of G0928+73.2600 could be reproduced by a planet with $M_{\rm p}\sin i$ as low as 2~$M_{\rm Jup}$  for an initial orbital separation of  1~AU.
 An accreted planet can contribute lithium to the star from its own stores, but  an expected upper limit to the planetary contribution is again the primordial lithium abundance of  3.28 dex. This limit can only be exceeded if  the planet has undergone chemical fractionation and is therefore enhanced in lithium itself.
To reach the lithium abundance seen in G0928+73.2600, we calculate that an accreted planet must have had  {\it A}(Li)$_{\rm p}=3.30+\log (1+M_{\rm env}/M_{\rm p})$. For 
a 2~$M_{\rm Jup}$ planet  and assuming 80\% of the stellar mass is in the envelope, this equation gives {\it A}(Li)$_{\rm p}=6.23$, which implies 850 times more lithium per hydrogen in the accreted object compared to our solar system's planets. 
A further test of this hypothesis would be to measure the abundance of other light elements, such as boron or beryllium.  These elements should  also be enriched if accreted planetary material  is responsible for the lithium enhancement.  
Searches for beryllium enhancements were carried out for a small number of lithium-rich stars \citep{cast99,melo05}, but no enhancements were found.

\section{Conclusion}
G0928+73.2600 joins the ranks of lithium-rich giant stars, and it may even be unique in this already unusual class.  The near-primordial lithium abundance suggests that either lithium depletion never began or, more likely, some replenishment mechanism has taken place.  However, this star's stellar parameters put it outside of 
the other groups of lithium-rich stars---the red bump stars or the  AGB stars---for which models of lithium regeneration exist.  G0928+73.2600's  evolutionary stage is consistent with either the base of
the RGB or the beginning of the AGB. It is unusual for lithium regeneration to have recently occurred at either phase.  The star's rotational velocity is higher than most red giant stars, and we
suggest that this could be explained by the accretion of a planet. Planet accretion may have triggered the lithium regeneration needed to explain the lithium abundance observed in the star.
Alternatively, the accretion of an {\it extremely} lithium-rich planet can account for the lithium enrichment of G0928+73.2600, its enhanced rotation, and  its pre-bump evolutionary stage.

\acknowledgments
J.K.C. appreciates support from the Virginia Space Grant Consortium and the NASA Earth and Space Science Fellowship, as well as
 travel support for observations from the National Optical Astronomy Observatory.
J.K.C. and S.R.M. acknowledge support from  the F.H. Levinson Fund of the Peninsula Community Foundation.
This study made use of VALD.

{\it Facilities:}  Mayall

\end{document}